\documentclass[12pt]{article}
\usepackage{amssymb}

\input{tcilatex}
\begin{document}

\bigskip \ 

\bigskip \ 

\begin{center}
\textbf{TOWARDS A CANONICAL GRAVITY IN TWO TIME}

\bigskip \textbf{\ }

\textbf{AND TWO SPACE DIMENSIONS}

\textbf{\ }

\smallskip \ 

J. A. Nieto\footnote{%
nieto@uas.uasnet.mx, janieto1@asu.edu}

\smallskip \ 

\textit{Facultad de Ciencias F\'{\i}sico-Matem\'{a}ticas de la Universidad
Aut\'{o}noma}

\textit{de Sinaloa, 80010, Culiac\'{a}n Sinaloa, M\'{e}xico}

\smallskip

and

\smallskip

\textit{Mathematical, Computational \& Modeling Sciences Center, Arizona
State University, PO Box 871904, Tempe AZ 85287, USA}

\bigskip \ 

\bigskip \ 

\bigskip \ 

\textbf{Abstract}
\end{center}

We describe a program for developing a canonical gravity in 2+2 dimensions
(two time and two space dimensions). Our procedure is similar to the usual
canonical gravity but with two times rather than just one time. Our work may
be of particular interest as an alternative approach to loop quantum gravity
in 2+2 dimensions.

\bigskip \ 

\bigskip \ 

\bigskip \ 

\bigskip \ 

\bigskip \ 


Pacs numbers: 04.60.-m, 04.65.+e, 11.15.-q, 11.30.Ly

November, 2011

\newpage

\noindent \textbf{1.- Introduction}

\smallskip \ 

It is well known that self-dual gravity [1]-[3] is one of the key concepts
in loop quantum gravity [4]. The general believe is that self-dual gravity
makes sense only in four dimensions since in this case the dual of a two
form (the curvature) is again a two form. However, there are a number of
evidences that self-dual gravity can also be implemented in eight dimensions
[5]-[8]. It turns out that even in four dimensions self-dual gravity does
not determine the signature of the `space-time'. In fact, it might be 1+3 or
0+4, as it is often considered in most of the current developments of loop
quantum gravity, but it might also be 2+2 as has been shown in Ref. [9],
where canonical gravity of the splitting type (1)+(1+2) was developed. It is
worth mentioning that a canonical approach with a splitting of the type
(1+1)+(2) has already be considered (see [10]-[12]and references therein).
However, these formalisms are still one time theory since refer to the $%
diag(-1,+1,+1,+1)$ signature rather than to $diag(-1,-1,+1,+1)$ signature.
In this work, we shall show that looking the scenario from the point of view
of two time physics one can also consider the splitting (2)+(2) (two time
and two space dimensions) of the `space-time'.

One of the main physical motivation for considering the splitting (2)+(2) of
the `space-time' comes from the possibility of finding a mechanism which may
allow to transform canonical gravity in 2+2 dimensions to canonical gravity
in 1+3 dimensions. This is equivalent to change one time-like dimension by
one space-like dimension and \textit{vice versa}. Surprisingly, this kind of
transformation has already be considered in the context of the sigma model
(see [13] and references therein) and one wonders whether similar map can be
implemented in canonical gravity changing 2+2 dimensions to 1+3 dimensions.
Here, we shall give a positive answer to such a question. Specifically, we
find an explicit evidence that the splitting of the action associated with
2+2 dimensions may lead to a term which has the typical form of a sigma
model action and therefore the transition from 2+2 dimensions to 1+3
dimensions (and \textit{vice versa}) is a viable possibility.

The contents of the paper are as follows: In section 2 and 3, the splittings
of the metric and the action associated with a 2+2 manifold are developed,
respectively. In section 4, we outline self dual gravity in 2+2 dimensions
with special emphasis of the group splitting $Spin(2,2)\sim SU(1,1)\times
SU(1,1)$. In section 5, we prove that, in a particular case, the action in
2+2 dimensions leads to a sigma model in which the usual method of changing
signature can be applied. Finally, in section 6 we make some final remarks
mentioning some topics of future interest.

\bigskip \ 

\noindent 2. \textbf{Splitting the metric associated with a }$2+2$\textbf{\
manifold}

\smallskip \ 

We shall assume that the vielbein field $E_{\hat{\mu}}^{~(\hat{A})}=E_{\hat{%
\mu}}^{~(\hat{A})}(t_{1},t_{2},x_{1},x_{2})=E_{\hat{\mu}}^{~(\hat{A})}(%
\mathbf{t},\mathbf{x})$ on a $2+2$-manifold $M^{2+2}$, can be written in the
form

\begin{equation}
E_{\hat{\mu}}^{~(\hat{A})}=\left( 
\begin{array}{cc}
e_{\mu }^{~(A)} & A_{\mu }^{~(a)} \\ 
B_{i}^{~(A)} & e_{i}^{~(a)}%
\end{array}%
\right) ,  \tag{1}
\end{equation}%
where $A_{\mu }^{~(a)}(\mathbf{t},\mathbf{x})\equiv E_{\mu }^{~(a)}(\mathbf{t%
},\mathbf{x})$ and $B_{i}^{~(A)}(\mathbf{t},\mathbf{x})\equiv E_{i}^{~(A)}(%
\mathbf{t},\mathbf{x})$. In (1), the notations $(\hat{A})$ and $\hat{\mu}$
of $E_{\hat{\mu}}^{~(\hat{A})}$ denote frame and target `space-time' indices
respectively. Of course, this form of $E_{\hat{\mu}}^{~(\hat{A})}$ resembles
a kind of Kaluza-Klein ansatz where one sets $B_{i}^{~(A)}=0$ [14]. The
inverse $E_{(\hat{A})}^{~~~\hat{\mu}}(\mathbf{t},\mathbf{x})$ can be
obtained from the relation 
\begin{equation}
E_{(\hat{A})}^{~~~\hat{\mu}}E_{\hat{\nu}}^{~(\hat{A})}=\delta _{\hat{\nu}}^{%
\hat{\mu}}.  \tag{2}
\end{equation}%
We find

\begin{equation}
E_{(\hat{A})}^{~~~\hat{\mu}}=\left( 
\begin{array}{cc}
e_{(A)}^{~~~\mu } & -A_{(A)}^{~~~i} \\ 
-B_{(a)}^{~~~\mu } & e_{(a)}^{~~~i}%
\end{array}%
\right) ,  \tag{3}
\end{equation}%
with 
\begin{equation}
A_{(A)}^{~~~i}\equiv e_{(a)}^{~~~i}A_{\mu }^{~(a)}e_{(A)}^{~~~\mu }  \tag{4}
\end{equation}%
and

\begin{equation}
B_{(a)}^{~~~\mu }\equiv e_{(A)}^{~~~\mu }B_{i}^{~(A)}e_{(a)}^{~~~i}.  \tag{5}
\end{equation}%
Here, we are assuming that%
\begin{equation}
e_{(A)}^{~~~\mu }e_{\nu }^{~(A)}=\delta _{\nu }^{\mu }  \tag{6}
\end{equation}%
and%
\begin{equation}
e_{(a)}^{~~~i}e_{j}^{~(a)}=\delta _{j}^{i}.  \tag{7}
\end{equation}%
Moreover, considering (6) and (7) one finds that (3) satisfies (2) provided
that the following relations are true:

\begin{equation}
e_{(A)}^{~~~\mu }B_{i}^{~(A)}e_{(a)}^{~~~i}A_{\nu }^{~(a)}=0  \tag{8}
\end{equation}%
and

\begin{equation}
e_{(a)}^{~~~i}A_{\mu }^{~(a)}e_{(A)}^{~~~\mu }B_{j}^{~(A)}=0.  \tag{9}
\end{equation}

Let $\eta _{(\hat{A}\hat{B})}$ be a flat $(2+2)$-metric. In general, the
metric $\gamma _{\hat{\mu}\hat{\nu}}$ can be defined in terms of $E_{\hat{\mu%
}}^{~(\hat{A})}$ in the usual form,

\begin{equation}
\gamma _{\hat{\mu}\hat{\nu}}=E_{\hat{\mu}}^{~(\hat{A})}E_{\hat{\mu}}^{~(\hat{%
B})}\eta _{(\hat{A}\hat{B})}.  \tag{10}
\end{equation}%
Using (1) the metric (10) becomes

\begin{equation}
\gamma _{\hat{\mu}\hat{\nu}}=\left( 
\begin{array}{cc}
e_{\mu }^{~(A)}e_{\nu }^{~(B)}\eta _{(AB)}+A_{\mu }^{~(a)}A_{\nu
}^{~(b)}\delta _{(ab)} & e_{\mu }^{~(A)}B_{j}^{~(B)}\eta _{(AB)}+A_{\mu
}^{~(a)}e_{j}^{~(b)}\delta _{(ab)} \\ 
B_{j}^{~(B)}e_{\nu }^{~(A)}\eta _{(AB)}+e_{j}^{~(b)}A_{\nu }^{~(a)}\delta
_{(ab)} & e_{i}^{~(a)}e_{j}^{~(b)}\delta
_{(ab)}+B_{i}^{~(A)}B_{j}^{~(B)}\eta _{(AB)}%
\end{array}%
\right) ,  \tag{11}
\end{equation}%
where $\eta _{(AB)}=diag(-1,-1)$, while $\delta _{(ab)}=diag(+1,+1)$. The
expression (11) can also be written as

\begin{equation}
\gamma _{\hat{\mu}\hat{\nu}}=\left( 
\begin{array}{cc}
g_{\mu \nu }+A_{\mu }^{~i}A_{\nu }^{~j}g_{ij} & g_{\mu \nu }B_{j}^{~\nu
}+g_{ij}A_{\mu }^{~j} \\ 
B_{j}^{~\mu }g_{\nu \mu }+g_{ji}A_{\nu }^{~i} & g_{ij}+B_{i}^{~\mu
}B_{j}^{~\nu }g_{\mu \nu }%
\end{array}%
\right) .  \tag{12}
\end{equation}%
Here, $g_{\mu \nu }=e_{\mu }^{~(A)}e_{\nu }^{~(B)}\eta _{(AB)}$, $%
g_{ij}=e_{i}^{~(a)}e_{j}^{~(b)}\delta _{(ab)},$ $A_{\mu
}^{~i}=e_{(a)}^{~~~i}A_{\mu }^{~(a)}$ and $B_{i}^{~\mu }=e_{(A)}^{~~~\mu
}B_{i}^{~(A)}$. Again if $B_{i}^{~(A)}=0$, (12) looks like a Kaluza-Klein
ansatz for the metric $\gamma _{\hat{\mu}\hat{\nu}}$.

Let $\Gamma _{\hat{\mu}\hat{\nu}}^{\hat{\alpha}}=\Gamma _{\hat{\nu}\hat{\mu}%
}^{\hat{\alpha}}$ and $\omega _{\hat{\mu}}^{~(\hat{A}\hat{B})}=-\omega _{%
\hat{\mu}}^{~(\hat{B}\hat{A})}$ be the Christoffel symbols and the spin
connection, respectively. We shall assume that $E_{\hat{\mu}}^{~(\hat{A})}$
satisfies the formula

\begin{equation}
\partial _{\hat{\mu}}E_{\hat{\nu}}^{~(\hat{A})}-\Gamma _{\hat{\mu}\hat{\nu}%
}^{\hat{\alpha}}E_{\hat{\alpha}}^{~(\hat{A})}+\omega _{\hat{\mu}}^{~(\hat{A}%
\hat{B})}E_{\hat{\nu}(\hat{B})}=0.  \tag{13}
\end{equation}
Using (13) it is not difficult to see that $\omega _{(\hat{A}\hat{B}\hat{C}%
)}=E_{(\hat{A})}^{~~~\hat{\mu}}\omega _{\hat{\mu}(\hat{B}\hat{C})}=-$ $%
\omega _{(\hat{A}\hat{C}\hat{B})}$ can be written in terms of%
\begin{equation}
\Omega _{\hat{\mu}\hat{\nu}}^{~~~(\hat{A})}=\partial _{\hat{\mu}}E_{\hat{\nu}%
}^{~(\hat{A})}-\partial _{\hat{\nu}}E_{\hat{\mu}}^{~(\hat{A})},  \tag{14}
\end{equation}%
in the form

\begin{equation}
\omega _{(\hat{A}\hat{B}\hat{C})}=\frac{1}{2}\left[ \Omega _{(\hat{A}\hat{B}%
\hat{C})}+\Omega _{(\hat{C}\hat{A}\hat{B})}-\Omega _{(\hat{B}\hat{C}\hat{A})}%
\right] .  \tag{15}
\end{equation}

\bigskip \ 

\bigskip \ 

\noindent 3. \textbf{Splitting the action associated with a }$2+2$\textbf{\
manifold}

\smallskip \ 

After some manipulation one can show that up to total derivative the action

\begin{equation}
S=\frac{1}{4}\int_{M^{2+2}}\sqrt{\gamma }R,  \tag{16}
\end{equation}%
is reduced to [14]%
\begin{equation}
S=-\frac{1}{4}\int_{M^{2+2}}E(\Omega _{(\hat{A}\hat{B}\hat{C})}\Omega ^{(%
\hat{A}\hat{B}\hat{C})}+2\Omega _{(\hat{A}\hat{B}\hat{C})}\Omega ^{(\hat{A}%
\hat{C}\hat{B})}-4\Omega _{(\hat{A}\hat{B})}^{~~~~~~(\hat{B})}\Omega
_{~~~~~~(\hat{C})}^{(\hat{A}\hat{C})}),  \tag{17}
\end{equation}%
where $E=\det E_{\hat{\mu}}^{~(\hat{A})}$. By using the splitting (1) one
may try to compute (16) via (17), but perhaps a simpler alternative may be
achieved by introducing the non-coordinate basis [15]

\begin{equation}
D_{\mu }=\partial _{\mu }-A_{\mu }^{i}\partial _{i}  \tag{18}
\end{equation}%
and

\begin{equation}
D_{i}=\partial _{i}-B_{i}^{\mu }\partial _{\mu }.  \tag{19}
\end{equation}%
The advantage of this basis is that brings the metric (11) into the block
diagonal form

\begin{equation}
\gamma _{\hat{\mu}\hat{\nu}}=\left( 
\begin{array}{cc}
g_{\mu \nu } & 0 \\ 
0 & g_{ij}%
\end{array}%
\right) .  \tag{20}
\end{equation}%
The case in which $B_{i}^{\mu }=0$ has already been considered by the
authors of the Ref [16]. They obtain that up to total derivative, the action
(16) becomes

\begin{equation}
\begin{array}{c}
S=\frac{1}{4}\int_{M^{2+2}}\sqrt{-\det g_{\mu \nu }}\sqrt{\det g_{ij}} \\ 
\times \{g^{\mu \nu }\hat{R}_{\mu \nu }+g^{ij}\tilde{R}_{ij}+\frac{1}{4}%
g_{ij}F_{\mu \nu }^{i}F^{\mu \nu j} \\ 
\\ 
+\frac{1}{4}g^{\mu \nu }g^{ij}g^{kl}[\mathcal{D}_{\mu }g_{ik}\mathcal{D}%
_{\nu }g_{jl}-\mathcal{D}_{\mu }g_{ij}\mathcal{D}_{\nu }g_{kl}] \\ 
\\ 
+\frac{1}{4}g^{ij}g^{\mu \nu }g^{\alpha \beta }[\partial _{i}g_{\mu \alpha
}\partial _{j}g_{\nu \beta }-\partial _{i}g_{\mu \nu }\partial _{j}g_{\alpha
\beta }]\},%
\end{array}
\tag{21}
\end{equation}%
(see Ref [16] for details). In the expression (21) the following definitions
are considered:

\begin{equation}
\hat{R}_{\mu \nu }=D_{\mu }\Gamma _{\alpha \nu }^{\alpha }-D_{\alpha }\Gamma
_{\mu \nu }^{\alpha }+\Gamma _{\mu \beta }^{\alpha }\Gamma _{\alpha \nu
}^{\beta }-\Gamma _{\beta \alpha }^{\beta }\Gamma _{\mu \nu }^{\alpha }, 
\tag{22}
\end{equation}%
\begin{equation}
\tilde{R}_{ij}=\partial _{i}\Gamma _{kj}^{k}-\partial _{k}\Gamma
_{ij}^{k}+\Gamma _{il}^{k}\Gamma _{kj}^{l}-\Gamma _{lk}^{l}\Gamma _{ij}^{k},
\tag{23}
\end{equation}%
\begin{equation}
F_{\mu \nu }^{i}=\partial _{\mu }A_{\nu }^{i}-\partial _{\nu }A_{\mu
}^{i}-A_{\mu }^{j}\partial _{j}A_{\nu }^{i}+A_{\nu }^{j}\partial _{j}A_{\mu
}^{i}  \tag{24}
\end{equation}%
and

\begin{equation}
\mathcal{D}_{\mu }g_{ij}=\partial _{\mu }g_{ij}-[A_{\mu }^{k}\partial
_{k}g_{ij}+\partial _{i}A_{\nu }^{k}g_{kj}+\partial _{j}A_{\nu }^{k}g_{ki}].
\tag{25}
\end{equation}

By symmetry, one may expect that in the most general case with $B_{i}^{\mu
}\neq 0$, the action%
\begin{equation}
\begin{array}{c}
S=\frac{1}{4}\int_{M^{2+2}}\sqrt{-\det g_{\mu \nu }}\sqrt{\det g_{ij}} \\ 
\times \{g^{\mu \nu }\hat{R}_{\mu \nu }+g^{ij}\tilde{R}_{ij}+\frac{1}{4}%
g_{ij}F_{\mu \nu }^{i}F^{\mu \nu j}+\frac{1}{4}g_{\mu \nu }H_{ij}^{\mu
}H^{\nu ij} \\ 
\\ 
+\frac{1}{4}g^{\mu \nu }g^{ij}g^{kl}[\mathcal{D}_{\mu }g_{ik}\mathcal{D}%
_{\nu }g_{jl}-\mathcal{D}_{\mu }g_{ij}\mathcal{D}_{\nu }g_{kl}] \\ 
\\ 
+\frac{1}{4}g^{ij}g^{\mu \nu }g^{\alpha \beta }[\mathcal{D}_{i}g_{\mu \alpha
}\mathcal{D}_{j}g_{\nu \beta }-\mathcal{D}_{i}g_{\mu \nu }\mathcal{D}%
_{j}g_{\alpha \beta }]\},%
\end{array}
\tag{26}
\end{equation}%
generalizes (21). Here, we used the definitions

\begin{equation}
H_{ij}^{\mu }=\partial _{i}B_{j}^{\mu }-\partial _{j}B_{i}^{\mu
}-B_{i}^{\alpha }\partial _{\alpha }B_{j}^{\mu }+B_{j}^{\alpha }\partial
_{\alpha }B_{i}^{\mu },  \tag{27}
\end{equation}%
\begin{equation}
\tilde{R}_{ij}=D_{i}\Gamma _{kj}^{k}-D_{k}\Gamma _{ij}^{k}+\Gamma
_{il}^{k}\Gamma _{kj}^{l}-\Gamma _{lk}^{l}\Gamma _{ij}^{k},  \tag{28}
\end{equation}%
and

\begin{equation}
\mathcal{D}_{i}g_{\mu \nu }=\partial _{i}g_{\mu \nu }-[B_{i}^{\alpha
}\partial _{\alpha }g_{\mu \nu }+\partial _{\mu }B_{i}^{\alpha }g_{\alpha
\nu }+\partial _{\nu }B_{i}^{\alpha }g_{\alpha \mu }].  \tag{29}
\end{equation}

In principle, as in Ref. [16] has been mentioned, the above method is
independent of the signature of the space-time. So, exactly the same result
can be obtained in the case of $m+n$-dimensional manifold which locally
looks like $M\times N$. In this context, the action (21) admits an
interpretation of a generally invariant gauge theory of $DiffN$ interacting
with gauged gravity and non-linear sigma field based on $M$. When $N$
corresponds to a group space $G$ the theory may admit an interpretation of
Kaluza-Klein type theory. In fact, in such a case one requires that $G$ is
an isometry of the $m+n$-dimensional metric and the resulting theory becomes
the Einstein-Yang-Mills-Sigma theory. In principle, in the case of the
generalized action (26) one can make a similar analysis. However, now one
needs to combine two possible interpretations. In fact, the action (26)
describes both a general invariant gauge theory of $DiffN$ based on $M$ and
a general invariant gauge theory of $DiffM$ based on $N$.

For our purpose it is convenient to recall that for $m=1$ and $n=3$ the
action (21) reduces to the canonical $1+3$ decomposition of four-dimensional
gravity. Since this scenario is generalized by Ashtekar formalism one
becomes motivated to look for a similar generalization for both (21) and
(26) actions. For $m=2$ and $n=2$, there are a number of works related to
(21), but not to (26). For instance, in Ref. [17] the self-dual Einstein
gravity is identified with $m=2$-dimensional sigma model with gauge symmetry 
$SDiffN^{2}$, the area preserving diffeomorphism of $N^{2}$. However, the
original signature of the metric is of the form $diag(-1,+1,+1,+1)$ rather
than $diag(-1,-1,+1,+1)$, as it is our interest in this work.

From the point of view of the signature $diag(-1,-1,+1,+1)$ there is not a
particular reason for assuming a $DiffN$ based on $M$ rather than $DiffM$
based on $N$. For this reason it is reasonable to consider the generalized
action (26) instead of (21). In this context, one observes that in addition
to the invariance of (26) under both $DiffN$ and $DiffM$, an immediate
symmetry of (26) is a kind of dual symmetry consist in the interchange of
both $g_{\mu \nu }\leftrightarrow g_{ij}$ and $A_{\mu }^{i}\leftrightarrow
B_{i}^{\mu }$. One can continue analyzing further properties of the action
(26), but here we are more interested in describing an outline for its
possible generalization in the context of Ashtekar formalism.

\bigskip \ 

\noindent 4. \textbf{Selfdual gravity in }$2+2$\textbf{\ dimensions}

\smallskip \ 

For our purpose, we recall that the self-dual curvature

\begin{equation}
^{+}R_{\hat{\mu}\hat{\nu}}^{(\hat{A}\hat{B})}=(R_{\hat{\mu}\hat{\nu}}^{(\hat{%
A}\hat{B})}+\frac{i}{2}\varepsilon _{~~~~(\hat{C}\hat{D})}^{(\hat{A}\hat{B}%
)}R_{\hat{\mu}\hat{\nu}}^{(\hat{C}\hat{D})})=-\frac{i}{2}\varepsilon _{~~~~(%
\hat{C}\hat{D})}^{(\hat{A}\hat{B})~~~~+}R_{\hat{\mu}\hat{\nu}}^{(\hat{C}\hat{%
D})},  \tag{30}
\end{equation}%
where $\varepsilon _{~~~~~~(\hat{C}\hat{D})}^{(\hat{A}\hat{B})}$ is the
completely antisymmetric density tensor, plays a central role in the
development of the Ashtekar formalism. Complex factor $i$ in (30) is linked
to the Lorenziana signature $diag(-1,+1,+1,+1)$. In the case of Euclidean
signature $diag(+1,+1,+1,+1)$ the imaginary factor $i$ in (30) can be
dropped:

\begin{equation}
^{+}R_{\hat{\mu}\hat{\nu}}^{(\hat{A}\hat{B})}=(R_{\hat{\mu}\hat{\nu}}^{(\hat{%
A}\hat{B})}+\frac{1}{2}\varepsilon _{~~~~(\hat{C}\hat{D})}^{(\hat{A}\hat{B}%
)}R_{\hat{\mu}\hat{\nu}}^{(\hat{C}\hat{D})})=\frac{1}{2}\varepsilon _{~~~~(%
\hat{C}\hat{D})}^{(\hat{A}\hat{B})~~~~+}R_{\hat{\mu}\hat{\nu}}^{(\hat{C}\hat{%
D})}.  \tag{31}
\end{equation}%
It turns out that in the signature $diag(-1,-1,+1,+1)$ one can also use
(31). Here, we would like to see what are the consequences of (31) in a
canonical approach. In the case of both Euclidean and Lorenziana signature,
(30) and (31) give $^{+}R_{\hat{\mu}\hat{\nu}}^{(ab)}=-i\varepsilon
_{~~~~(c)}^{(ab)~~+}R_{\hat{\mu}\hat{\nu}}^{(c0)}$ and $^{+}R_{\hat{\mu}\hat{%
\nu}}^{(ab)}=\varepsilon _{~~~~(c)}^{(ab)~~+}R_{\hat{\mu}\hat{\nu}}^{(c0)}$
respectively and therefore one observes that in both cases the $^{+}R_{\hat{%
\mu}\hat{\nu}}^{(cd)}$ component can be written in terms of $^{+}R_{\hat{\mu}%
\hat{\nu}}^{(a0)}$. Moreover, symbolically one has $^{+}R_{\hat{\mu}\hat{\nu}%
}^{(a0)}\sim \partial _{\hat{\mu}}^{~+}\omega _{\hat{\nu}}^{(a0)}-\partial _{%
\hat{\nu}}^{~+}\omega _{\hat{\mu}}^{(a0)}+...$ and thus one can consider $F_{%
\hat{\mu}\hat{\nu}}^{a}\equiv $ $^{+}R_{\hat{\mu}\hat{\nu}}^{(a0)}$ as the
Yang-Mills field strength and $A_{\hat{\mu}}^{a}=$ $^{+}\omega _{\hat{\mu}%
}^{(a0)}$ as the gauge field, with $SU(2)$ as a gauge group. Roughly
speaking, these observations are some of the key reasons behind the success
of the Ashtekar formalism. However, in the case of the signature $%
diag(-1,-1,+1,+1)$ the scenario seems to be different. This is because in
such case there is not a particular reason for considering the splitting of
(31) in terms of only one time coordinate (see Ref. [9]) instead of two
times coordinates. Specifically, in this case one has the following
splitting of (31):

\begin{equation}
^{+}R_{\hat{\mu}\hat{\nu}}^{(AB)}=\frac{1}{2}\varepsilon
_{~~~~(cd)}^{(AB)~~~+}R_{\hat{\mu}\hat{\nu}}^{(cd)}=\frac{1}{2}\varepsilon
^{(AB)}\varepsilon _{(cd)}\text{ }^{+}R_{\hat{\mu}\hat{\nu}}^{(cd)}, 
\tag{32}
\end{equation}

\begin{equation}
^{+}R_{\hat{\mu}\hat{\nu}}^{(Aa)}=\varepsilon _{~~~~(Bb)}^{(Aa)~~~~+}R_{\hat{%
\mu}\hat{\nu}}^{(Bb)}=\varepsilon _{~~~~(B)}^{(A)}\varepsilon
_{~~~~(b)}^{(a)~~~~+}R_{\hat{\mu}\hat{\nu}}^{(Bb)},  \tag{33}
\end{equation}%
and%
\begin{equation}
^{+}R_{\hat{\mu}\hat{\nu}}^{(ab)}=\varepsilon _{~~~~(AB)}^{(ab)~~~~~+}R_{%
\hat{\mu}\hat{\nu}}^{(AB)}=\varepsilon ^{(ab)}\varepsilon _{(AB)}\text{ }%
^{+}R_{\hat{\mu}\hat{\nu}}^{(AB)}.  \tag{34}
\end{equation}%
Clearly, (32) and (34) are equivalent expressions. The formula (33) simply
seems a indices relation, between the different components of the frame
indices of the object $^{+}R_{\hat{\mu}\hat{\nu}}^{(Aa)}$. However, one can
verify that (33) reduces the four frame indices components of $^{+}R_{\hat{%
\mu}\hat{\nu}}^{(Aa)}$ to only two independent components. Finally, one
notes that (32) determines $^{+}R_{\hat{\mu}\hat{\nu}}^{(AB)}$ in terms of $%
^{+}R_{\hat{\mu}\hat{\nu}}^{(cd)}$ and vice versa. But in two dimensions one
can write $^{+}R_{\hat{\mu}\hat{\nu}}^{(AB)}=\varepsilon ^{(AB)~+}R_{\hat{\mu%
}\hat{\nu}}$, where $^{+}R_{\hat{\mu}\hat{\nu}}=\frac{1}{2}\varepsilon
_{(CD)}^{~~~~+}R_{\hat{\nu}}^{CD}$. So, symbolically, in this case, one
expects to have $^{+}R_{\hat{\mu}\hat{\nu}}\sim \partial _{\hat{\mu}%
}^{~+}\omega _{\hat{\nu}}-\partial _{\hat{\nu}}^{~+}\omega _{\hat{\mu}}$,
where $\omega _{\hat{\mu}}=\frac{1}{2}\varepsilon _{(CD)}$ $^{+}\omega _{%
\hat{\mu}}^{(CD)}$, and therefore $^{+}\omega _{\hat{\nu}}$ can be
understood as an Abelian gauge field. Similarly, we can write $^{+}R_{\hat{%
\mu}\hat{\nu}}^{(Aa)}\sim \partial _{\hat{\mu}}^{~+}\omega _{\hat{\nu}%
}^{(Aa)}-\partial _{\hat{\mu}}^{~+}\omega _{\hat{\nu}}^{(Aa)}+...$, with $%
^{+}\omega _{\hat{\nu}}^{(Aa)}$ corresponding to only two additional
independents gauge fields.

It may be helpful to analyze the above scenario from the point of view of
group splitting. In general the splitting of the curvature can be related to
the splitting of the connection. In turn the splitting of the connection is
related to group algebra splitting. In the case of Euclidean signature the
splitting of the curvature $R_{\hat{\mu}\hat{\nu}}^{(\hat{A}\hat{B})}$ in
terms of self-antiself dual parts has its origins in the splitting $%
Spin(4)\sim Spin(3)\times Spin(3)$, while in the case of Lorenzian signature
one has $SO(1,3)\sim SU(2)\times SU(2)$ (see Refs. [18]-[20]). In Ref. [9]
is mentioned that in an scenario of $2+2$ dimensions one may consider the
splitting $SO(2,2)\sim SL(2,R)\times SL(2,R)$. This observation may in
principle be extended to an splitting of the form $Spin(2,2)\sim
SU(1,1)\times SU(1,1)$. This is because there exist the isomorphism $%
SL(2,R)\sim SU(1,1)$. However, one should mention that these kind of
splittings are not sufficient for the a consistent splitting of the
curvature. In fact, one still needs to verify that at the level of the
corresponding algebra the self-dual $^{+}\omega $ and antiself-dual $%
^{-}\omega $ parts of the connection $\omega $ are in fact connections of
the corresponding group: $SL(2,R)$ or $SU(1,1)$ in our case. This can be
accomplish by splitting the $SU(2,2)$ gauge transformation into two $SU(1,1)$
gauge transformations and checking that the self-dual and antiself dual
connections behave properly under the reduced gauge transformations
associated with $SU(1,1)$.

\bigskip \ 

\noindent 5. \textbf{From canonical gravity in }$2+2$ to $1+3$ \textbf{%
dimensions}

\smallskip \ 

Here, we shall give an outline of the possibility to apply a map to the
action (26) in such a way that one can go from gravity in $2+2$ dimensions
to gravity in $1+3$ dimensions. Our mechanism is similar to the one used in
a sigma model theory (see Ref. [13] and references therein).

Let us start by recalling how starting from the action (21) one can obtain
the usual canonical gravity in $1+3$ dimensions. In this case one obtains
exactly the same action as (21) but with $M^{1+3}$ instead of $M^{2+2}$.
Thus, we take the index $m=1$ and the index $n=2,3$ and $4$. One discovers
that the action (21) is reduced to

\begin{equation}
S=\frac{1}{4}\int_{M^{1+3}}\sqrt{\det (g_{ij})}\{g^{ij}\tilde{R}_{ij}++\frac{%
1}{4}g^{\mu \nu }g^{ij}g^{kl}[\mathcal{D}_{\mu }g_{ik}\mathcal{D}_{\nu
}g_{jl}-\mathcal{D}_{\mu }g_{ij}\mathcal{D}_{\nu }g_{kl}]\},  \tag{35}
\end{equation}%
with

\begin{equation}
\mathcal{D}_{1}g_{ij}=\partial _{1}g_{ij}-[A_{1}^{k}\partial
_{k}g_{ij}+\partial _{i}A_{1}^{k}g_{kj}+\partial _{j}A_{1}^{k}g_{ki}]. 
\tag{36}
\end{equation}%
Defining the extrinsic curvature as

\begin{equation}
K_{ij}\equiv \mathcal{D}_{1}g_{ij},  \tag{37}
\end{equation}%
one sees that (35) can be rewritten in the form

\begin{equation}
S=\frac{1}{4}\int_{M^{1+3}}\sqrt{\det (g_{ij})}\{g^{ij}\tilde{R}_{ij}+\frac{1%
}{4}g^{ij}g^{kl}[K_{ik}K_{jl}-K_{ij}K_{kl}]\},  \tag{38}
\end{equation}%
which is the typical action for canonical gravity in $1+3$ dimensions.

In view of the above review we see that besides the curvature term $g^{ij}%
\tilde{R}_{ij}$ the relevant term is the second term in (35). For this
reason we shall focus in the term:

\begin{equation}
S=\frac{1}{16}\int_{M^{2+2}}\sqrt{\det (g_{ij})}g^{\mu \nu }g^{ij}g^{kl}[%
\mathcal{D}_{\mu }g_{ik}\mathcal{D}_{\nu }g_{jl}-\mathcal{D}_{\mu }g_{ij}%
\mathcal{D}_{\nu }g_{kl}].  \tag{39}
\end{equation}%
Four our purpose, we shall take $\mathcal{D}_{\mu }g_{ik}=\partial _{\mu
}g_{ik}$. Moreover, we shall call $\Phi ^{(p)}$ with $(p)=1,2,3$ the three
degrees of freedom associated with the the two dimensional metric $g_{ik}$.
So the action (39) becomes

\begin{equation}
\begin{array}{c}
S=\frac{1}{16}\int_{M^{2+2}}\sqrt{\det (g_{\alpha \beta })}g^{\mu \nu
}\partial _{\mu }\Phi ^{(p)}\partial _{\mu }\Phi ^{(q)}h_{(pq)},%
\end{array}
\tag{40}
\end{equation}%
where

\begin{equation}
h_{(pq)}=g^{ij}g^{kl}(\frac{\partial g_{ik}}{\partial \Phi ^{(p)}}\frac{%
\partial g_{jl}}{\partial \Phi ^{(q)}}-\frac{\partial g_{ij}}{\partial \Phi
^{(p)}}\frac{\partial g_{kl}}{\partial \Phi ^{(q)}}).  \tag{41}
\end{equation}%
We recognize in (40) a sigma model type action. Since in principle, $g_{ik}$
is different when one is considering a theory with $2+2$ signature or with $%
1+3$ signature, one can consider the fact that the metric $h_{pq}$ is
signature dependent. To illustrate how duality may work by starting with the
action (40) we shall further simplify the scenario. Let us assume that $%
g_{\mu \nu }$ and $h_{(pq)}$ are flat metrics $\delta _{\mu \nu }$ and $\eta
_{(pq)}$ respectively. So, (40) becomes

\begin{equation}
\begin{array}{c}
S=\frac{1}{16}\int_{M^{2+2}}\delta ^{\mu \nu }\partial _{\mu }\Phi
^{(p)}\partial _{\nu }\Phi ^{(q)}\eta _{(pq)}.%
\end{array}
\tag{42}
\end{equation}%
We shall also assume that $\delta _{\mu \nu }$ refers to the Euclidean
sector of both $2+2$ and $1+3$ signatures. This means that $\eta _{(pq)}$
will depends on the two times associated with $2+2$ signature, or one time
and one space in the case of the $1+3$ signature. In other words, we shall
assume that in the case of $2+2$ signature, $\eta _{(pq)}$ takes the form $%
\eta _{(pq)}=diag(1,1,1)$, while in the sector of $1+3$ signature $\eta
_{(pq)}$ is given by $\eta _{(pq)}=diag(-1,1,1)$ . Thus, our task is to see
how one can go from $\eta _{(11)}=1$ in the case of $2+2$ dimensions to $%
\eta _{(11)}=-1$ in the case of $1+3$ dimensions. Therefore we focus in the
reduced action

\begin{equation}
\begin{array}{c}
S=\frac{1}{16}\int_{M^{2+2}}\delta ^{\mu \nu }\partial _{\mu }\Phi
^{(1)}\partial _{\nu }\Phi ^{(1)}\eta _{11}=\frac{1}{16}\int_{M^{2+2}}\delta
^{\mu \nu }\partial _{\mu }\Phi ^{(1)}\partial _{\nu }\Phi ^{(1)}.%
\end{array}
\tag{43}
\end{equation}%
The next step is a standard procedure. We introduce an auxiliary gauge field 
$\mathcal{A}_{\mu }$ and add to (43) a term $\varepsilon ^{\mu \nu }\mathcal{%
A}_{\mu }\partial _{\nu }\Psi ^{(1)}$, where $\Psi ^{(1)}$ is a dual field.
Thus, (43) becomes

\begin{equation}
\begin{array}{c}
S=\frac{1}{8}\int_{M^{2+2}}\frac{1}{2}\delta ^{\mu \nu }(\partial _{\mu
}\Phi ^{(1)}-\mathcal{A}_{\mu })(\partial _{\nu }\Phi ^{(1)}-\mathcal{A}%
_{\nu })+\varepsilon ^{\mu \nu }\mathcal{A}_{\mu }\partial _{\nu }\Psi
^{(1)}.%
\end{array}
\tag{44}
\end{equation}%
The symmetries of the theory allows us to set $\mathcal{A}_{\mu }=0$ or $%
\Phi ^{(1)}=0$. In the first case the action (44) is reduced to (43). While
in the second case by setting $\Phi ^{1}=0$ in (44) one gets

\begin{equation}
\begin{array}{c}
S=\frac{1}{8}\int_{M^{2+2}}\frac{1}{2}\delta ^{\mu \nu }\mathcal{A}_{\mu }%
\mathcal{A}_{\nu }+\varepsilon ^{\mu \nu }\mathcal{A}_{\mu }\partial _{\nu
}\Psi ^{(1)}.%
\end{array}
\tag{45}
\end{equation}%
Solving (45) for $\mathcal{A}_{\mu }$, one obtains%
\begin{equation}
\mathcal{A}^{\mu }+\varepsilon ^{\mu \nu }\partial _{\nu }\Psi ^{(1)}=0. 
\tag{46}
\end{equation}%
Substituting this result into (45) yields the dual action

\begin{equation}
\begin{array}{c}
S=\frac{1}{16}\int_{M^{1+3}}(-1)\delta ^{\mu \nu }\partial _{\mu }\Psi
^{(1)}\partial _{\nu }\Psi ^{(1)}.%
\end{array}
\tag{47}
\end{equation}%
The minus sign in (47) means that we have be able to change the value of $%
\eta _{11}$ from $1$ to $-1$ as expected. In turn this means that the
original flat metric $\eta _{(pq)}=diag(1,1,1)$ corresponding to the $2+2$
signature becomes, in the dual sector, the flat metric $\eta
_{(pq)}=diag(-1,1,1)$ associated with the $1+3$ signature. Presumably, this
procedure can be, of course, generalized for non-flat metrics $g_{\mu \nu }$
and $h_{(pq)}$, but this will require some additional computations.

It turns out that signature changes can be connected with topology changes
[21]. So, it may be interesting to relate our present procedure of signature
change with that of topological change.

\bigskip \ 

\noindent 6. \textbf{Final remarks}

\smallskip \ 

Summarizing we have described a self-dual gravitational theory in which the
signature corresponds to two time and two space dimensions, that is to the
signature $diag(-1,-1,+1,+1)$. Our preliminary analysis indicates that an
action of the form 
\begin{equation}
S=\frac{1}{4}\int_{M^{2+2}}\text{ }E\text{ }E_{(\hat{A})}^{~\hat{\mu}}E_{(%
\hat{B})}^{~\hat{\nu}}\text{ }^{+}R_{\hat{\mu}\hat{\nu}}^{(\hat{A}\hat{B})},
\tag{48}
\end{equation}%
will describe a self-dual gravitational gauge theory with a gauge field with
only three degrees of freedom. Of course, in order to have a complete theory
one needs to develop (48) in full details, but in this sense our proposed
action (26) surely may provide an important mathematical tool for such a
purpose.

Finally, it is worth mentioning that one of the main motivations in Ref. [9]
was the idea of establishing a connection between Ashtekar formalism in $%
diag(-1,-1,+1+1)$ signature and oriented matroid matroid theory [22] (see
also Ref. [23]-[24] and references therein). We believe that the present
work can be also useful in such a quest.

\bigskip \ 

Note added: While we were preparing this paper we became aware of the Refs.
[25]-[26], where new variables for classical and quantum gravity in higher
dimensions are discussed. It will be interesting for further research to see
whether there is a connection between the present work and such references.

\bigskip \ 

\noindent \textbf{Acknowledgments: }I would like to thank the referees for
helpful comments. This work was partially supported by PROFAPI-2011.

\smallskip \

\end{document}